\documentclass[12pt,a4paper]{article}

\usepackage{jheppub}
\usepackage{comment}
\usepackage{graphicx}
\usepackage{amsmath}
\usepackage{dsfont}
\usepackage{slashed}
\usepackage{braket}
\usepackage{color}

\global\long\def\D{\Delta}

\global\long\def\p{\partial}

\global\long\def\a{\alpha}
\global\long\def\b{\beta}

\global\long\def\g{\gamma}

\global\long\def\l{\lambda}

\global\long\def\e{\epsilon}
\global\long\def\r{\rho}

\newcommand{\cO}{\ensuremath{\mathcal{O}}}

\newcommand{\cK}{\ensuremath{\mathcal{K}}}
\newcommand{\dr}{_{,\rho}}

\newcommand{\dho}{\partial}

\newcommand{\BY}{\textsc{by}}
\newcommand{\FT}{\textsc{ft}} % small-caps
\newcommand{\scalar}{\text{scalar}}
\newcommand{\pure}{\text{pure}}
\newcommand{\bulk}{\mathrm{bulk}}

\DeclareMathOperator{\trace}{Tr}

\newcommand{\ed}{\,.}
\newcommand{\ec}{\,,}
\newcommand{\ecq}{\,,\quad}
\newcommand{\nn}{\nonumber}

\newcommand{\theeta}{\eta}

\newcommand{\comments}[1]{}

\title{The Holographic Dictionary for Beta Functions of Multi-trace Coupling Constants}

\author{
Ofer Aharony$^a$, Guy Gur-Ari$^{a,b}$ and Nizan Klinghoffer$^a$
\\
\it{$^a$ Department of Particle Physics and Astrophysics,\\
Weizmann Institute of Science, Rehovot 7610001, Israel}
\\ 
\it{$^b$ Stanford Institute for Theoretical Physics, \\
Stanford University, Stanford, California 94305, USA
}
}
\emailAdd{Ofer.Aharony@weizmann.ac.il}
\emailAdd{guyga@stanford.edu} 
\emailAdd{Nizan.Klinghoffer@weizmann.ac.il}

\abstract{
Field theories with weakly coupled holographic duals, such as large $N$ gauge theories, have a natural separation of their operators into `single-trace operators' (dual to single-particle states) and `multi-trace operators' (dual to multi-particle states).
There are examples of large $N$ gauge theories where the beta functions of single-trace coupling constants all vanish, but marginal multi-trace coupling constants have non-vanishing beta functions that spoil conformal invariance (even when all multi-trace coupling constants vanish).
The holographic dual of such theories should be a classical solution in anti-de Sitter space, in which the boundary conditions that correspond to the multi-trace coupling constants depend on the cutoff scale, in a way that spoils conformal invariance.
We argue that this is realized through specific bulk coupling constants that lead to a running of the multi-trace coupling constants.
This fills a missing entry in the holographic dictionary.
}

\begin{document}
\maketitle
 \bibliographystyle{JHEP}

\section{Introduction and Summary of Results}

The gauge/gravity duality relates gauge theories to gravitational theories, and in particular it relates conformal field theories (CFTs) in $d$ space-time dimensions to gravitational theories on $(d+1)$-dimensional anti-de Sitter ($AdS$) space \cite{Maldacena:1997re}.
This correspondence is most useful when the gravitational theory is weakly coupled, at least at low energies; this situation arises in particular for gauge theories in the 't~Hooft large $N$ limit, for which the couplings in the gravitational theory scale as $1/N$.
In such a case (which we will assume throughout this paper) we can use a classical approximation for the gravitational theory.
On the gravity side there is then a clear separation between single-particle and multi-particle states, with small mixing between them.
States on the gravitational side map to operators in the dual field theory.
Single-particle states map to operators called `single-trace operators', since in the large $N$ gauge theory example, such operators arise as single traces of products of fields in the adjoint representation.
Multi-particle states map to `multi-trace operators'; in the limit where the gravity theory is weakly coupled, these operators can be thought of as the non-singular terms in the product of several `single-trace operators' that are taken to the same point.

In the original formulation of gauge/gravity duality, only field theory actions involving single-trace operators were considered. It was understood \cite{Gubser:1998bc,Witten:1998qj} how to relate the coupling constants for these operators to the boundary conditions of the corresponding bulk fields (that create the single-particle states). Marginal coupling constants are particularly interesting, since when they are exactly marginal, turning them on can give families of conformal field theories, while if they have non-trivial beta functions then conformal invariance is lost. In the bulk theory, marginal single-trace operators map to massless bulk scalar fields, and turning on coupling constants for these operators corresponds to giving a vacuum expectation value to these scalar fields. A single-trace operator is exactly marginal if there is a solution preserving the isometries of $AdS$ space for every value of the corresponding bulk scalar field, while otherwise the operator has a non-zero beta function; examples of both types exist already in the simple example of the $d=4$, ${\cal N}=4$ supersymmetric Yang-Mills (SYM) theory, see for instance \cite{Aharony:2002hx}. The beta functions for these operators can be related to the bulk solutions that arise when turning on the corresponding scalar fields.

Multi-trace operators are usually irrelevant, since their dimension (in the classical gravity limit) is the sum of the dimensions of the single-trace operators that they are made of. However, there are cases when such operators can be relevant or marginal. In the context of the gauge/gravity duality, turning on coupling constants for these operators was first discussed in \cite{Aharony:2001pa} from the point of view of the dual string worldsheet theory, and a simple bulk picture for the corresponding coupling constants was then provided in \cite{Witten:2001ua,Berkooz:2002ug}. In this picture these coupling constants are related to non-linear boundary conditions for the bulk fields that are dual to the single-trace operators making up the multi-trace operator.
These boundary conditions do not affect the classical solution in the absence of sources, but they affect the fluctuations around it.

When we have marginal multi-trace operators, it is interesting to ask if they are exactly marginal or if they have a non-zero beta function. Such a beta function in general depends (even in the large $N$ limit) both on the single-trace coupling constants and on the multi-trace coupling constants. Examples of exactly marginal multi-trace operators were presented for instance in \cite{Aharony:2001pa,Witten:2001ua}, while a simple example of a non-zero beta function for a double-trace operator was analyzed in \cite{Witten:2001ua} (and is reviewed below).

A particularly interesting situation is when we have a field theory in which all single-trace beta functions vanish. Naively one would assume that such a theory must be conformal (at least in the large $N$ limit). However, as analyzed for instance in \cite{Dymarsky:2005uh,Dymarsky:2005nc,Pomoni:2008de}, such theories could still have (even in the large $N$ limit, and even when all multi-trace coupling constants vanish) beta functions for multi-trace coupling constants that would spoil conformal invariance. In particular, this is the typical situation at weak coupling in theories that arise from `orbifolds' of the ${\cal N}=4$ SYM theory (by which we mean the field theories dual to orbifolds of the corresponding string theory \cite{Kachru:1998ys,Lawrence:1998ja,Bershadsky:1998cb}). It was shown in \cite{Dymarsky:2005uh,Dymarsky:2005nc,Pomoni:2008de} that at weak coupling these theories have marginal multi-trace operators arising from products of single-trace operators in the `twisted sector' (that do not directly inherit their properties from the ${\cal N}~=~4$ SYM theory), and that in non-supersymmetric orbifolds at weak coupling these multi-trace operators always have non-zero beta functions (even when the multi-trace coupling constants vanish). Moreover it was found that there is no weakly coupled solution to the beta function equations, so that conformal invariance is broken in these field theories. This precludes using these orbifolds as simple non-supersymmetric examples of the $AdS$/CFT correspondence. Similar results were obtained for specific non-supersymmetric theories that arise as deformations of the $\mathcal{N}=4$ SYM theory \cite{Fokken:2013aea}.

The main question we would like to answer in this paper is how such a situation is reflected in the dual bulk physics. As described above, having a vanishing beta function for all single-trace couplings maps to having a classical gravitational solution in $AdS$ space. Naively one would expect this to be enough to ensure that the dual field theory is conformal, but the discussion above implies that even in such a situation there could be beta functions for multi-trace operators that would spoil conformal invariance. How are such beta functions realized in the $AdS$/CFT correspondence~? We expect to find that when we perform a holographic renormalization on $AdS$ space (as we usually need to do in order to carefully cancel all the divergences in classical bulk computations), the value of the multi-trace coupling constants will depend on the renormalization scale, so that we cannot set them to zero at all scales, and that this will break conformal invariance (despite the fact that the classical bulk solution in the absence of sources is conformally invariant).

We will show that indeed this happens, and that when we have marginal multi-trace operators, there are specific types of coupling constants in the bulk that are related to the beta functions for the multi-trace operators. For simplicity we work at leading order in these coupling constants, and we show that their presence leads to a scale-dependence of the multi-trace coupling constants which does not allow setting them to zero at all scales. We emphasize the contribution to multi-trace beta functions that persists even when the multi-trace couplings vanish (namely, it is a function just of the single-trace couplings), since other contributions that depend on the multi-trace couplings are more straightforward (they are often determined just by conformal perturbation theory) and were analyzed already in the literature.

In principle, one example in which this situation should arise is the orbifolds of ${\cal N}=4$ SYM discussed above. However, that situation is at weak 't Hooft coupling in the field theory, so its bulk dual description is complicated, involving higher-derivative interactions and light higher-spin fields (though the bulk is still weakly coupled in the large $N$ limit). 
In the same theories at strong coupling, the multi-trace operators are no longer marginal, since (at least when the orbifold has no fixed points on the $S^5$) the single-trace operators that they are made of have large anomalous dimensions, so this issue does not arise. But we expect that there should be many other situations where such beta functions could be important (in particular with little or no supersymmetry).

Beta functions for double-trace operators were previously discussed in \cite{Vecchi:2010dd}; in that paper a different suggestion was given for the bulk mapping of the beta functions, which seems not to agree with ours. 
%Beta functions for multi-trace operators were also discussed in \cite{Das:2013qea}; they focused on different issues than we do, but as far as we can see their results agree with ours wherever they overlap.
Beta functions for multi-trace operators were previously discussed in \cite{Das:2013qea},
which focused on different issues than we do, and in \cite{Grozdanov:2011aa} in the
somewhat different framework of the Wilsonian holographic renormalization group.
As far as we can see the results of \cite{Das:2013qea,Grozdanov:2011aa} agree with ours whenever
they overlap.

We begin in section \ref{two} with a review of the holographic dictionary for multi-trace couplings and for computing their beta functions. In section \ref{suggestion} we give an argument for how a beta function for multi-trace couplings that depends on single-trace couplings should show up in the holographic dual, and we then test this suggestion for multi-trace couplings involving three or more operators in section \ref{multi-trace}. In section
\ref{doubletrace1} we discuss the special case of double-trace operators. Finally, in section \ref{stress} we present an alternative computation of the beta function, using the expectation value of the trace of the stress tensor. Appendices contain some technical results.

\section{A Review of Multi-trace Deformations and Holography}
\label{two}

In this section we review the holographic dictionary for multi-trace coupling constants, which are related to boundary conditions of the dual fields.
For simplicity we will use the language of large $N$ gauge theories, though everything we say can be generalized to other theories that have
weakly coupled bulk duals.

Consider a gauge theory in $d$ space-time dimensions with adjoint-valued fields $\Phi$, and a set of single-trace operators
\begin{align}
  \cO_i = \frac{1}{N} \trace (f_i(\Phi)) \ed
\end{align}
Here, $f_i$ are polynomial functions of the fields $\Phi$ and their derivatives that do not depend explicitly on $N$.
The action for this theory is taken to be
\begin{align}
  S = - N^2 W(\cO_i) \ec
  \label{SW}
\end{align}
where $W$ does not depend explicitly on $N$.
We define the 't Hooft large $N$ limit by taking $N \to \infty$ while keeping fixed the couplings that define $W$.
With these conventions, it is easy to see that the planar 1-point functions $\langle \cO_i \rangle$ are $O(N^0)$,\footnote{
Notice that the $\Phi$ propagator is $O(N^{-1})$ in this case.
In a model with vector-valued fields $\hat{\Phi}$ we would define the `single-trace' operators to be $\hat{\cO}_i = N^{-1} \hat{\Phi} \cdot g_i(\overleftarrow{\dho},\overrightarrow{\dho}) \hat{\Phi}$, and the action would be given by $S = - N {\hat W}(\hat{\cO}_i)$.
In the large $N$ limit we would keep $\hat W$ fixed, and the planar 1-point functions would again be independent of $N$.
}
and we can directly identify each 1-point function with a mode of the holographically dual field (when the bulk
fields are normalized such that the bulk action is also proportional to $N^2$; for simplicity we will drop this
overall normalization in the rest of this paper).

When the field theory is conformal, its holographic dual is a theory of gravity on $AdS_{d+1}$.  We choose
the background metric on the Poincar\'e patch of $AdS_{d+1}$ to be
\begin{align}
  ds^2 = \frac{dz^2 + dx_i dx^i}{z^2} \ec
  \label{metric}
\end{align}
with the boundary at $z=0$, and we set the $AdS$ radius to 1.\footnote{We will use Greek letters $\mu, \nu, \cdots$ to denote coordinates on the full $AdS$ and Roman letters $i,j,\cdots$ to denote coordinates on the boundary of $AdS$.}

Consider a single-trace scalar operator $\cO$ of dimension 
\begin{align}
  \D = \frac{d}{2} - \nu \ecq 0 < \nu < 1 \ed
\end{align}
We choose the range $\D < d/2$ because we are interested in cases where multi-trace deformations that involve $\cO$ are marginal
(we will separately discuss the double-trace case below).
The lower bound on $\Delta$ is due to unitarity.
This operator is holographically dual to a scalar field $\phi$ with squared mass $m^2 = \Delta (\Delta - d)$.
Near the boundary, the scalar field has the mode expansion
\begin{align}
\phi(x,z) = \a(x) z^{\D} + \b(x)z^{d-\D} + \cdots \ed
\label{phimodeexp}
\end{align}
The boundary conditions of $\phi$ at $z=0$ are a relation between $\beta(x)$ and $\alpha(x)$, that is given by\footnote{
Note that our conventions differ from those of \cite{Witten:2001ua} by a minus sign, due to our definition of $W$.
}
\cite{Witten:2001ua}
\begin{align}
\label{bcdef}
  \beta(x) = \left. \frac{\delta W}{\delta \cO(x)} \right|_{\cO(y)=-2\nu\a(y)} \ed
\end{align}
When there are several operators $\cO_i$, the boundary conditions are given by
\begin{align}
  \beta_i = 
  \left. \frac{\delta W}{\delta \cO_i} \right|_{\cO_j = -2\nu_j\alpha_j}
  \ed
  \label{manyBC}
\end{align}
The precise normalization is due to the fact that the expectation value of the operator is related to the bulk scalar \eqref{phimodeexp} by $\langle \cO(x) \rangle = -2\nu\alpha(x)$, when using the ordinary holographic prescription \cite{Klebanov:1999tb}.
We rederive this relation in appendix \ref{free}.

For example, suppose that $W$ depends on $\cO$ through the multi-trace interaction $W = \lambda_\FT \int \! d^dx \, \cO^n / n$.
The boundary conditions for the dual scalar are then given by
$\beta = \lambda_\FT (-2\nu\alpha)^{n-1}$.
When all multi-trace coupling constants vanish, $W$ is linear in $\cO$ so that $\b$ is fixed by the boundary conditions and is identified with the source of $\cO$, while $\a$ is undetermined by these conditions and it is related to the expectation value of $\cO$.

\subsection{Example: Double-trace Beta Function}
\label{sec:DoubleTrace}

In this section we review the holographic calculation of the beta function of a marginal double-trace operator \cite{Witten:2001ua}.
Consider a scalar operator $\cO$ of dimension $\D = d/2$.
The dual scalar field in this case has the near-boundary expansion
\begin{align}
  \phi(x,z) = \alpha(x) z^{d/2} + \beta(x) z^{d/2} \log(\mu z) + \cdots \ec
\end{align}
where we interpret $\mu$ as a renormalization scale.
Here $\beta$ corresponds to the source of $\cO$, while $\alpha$ is related to its expectation value.

Let us turn on a double-trace deformation $W = -\lambda_\FT \int \! d^dx \, \cO^2 / 4$.
In this case the relation between the VEV and the bulk mode is $\langle \cO \rangle = -2\alpha$ (this is derived in appendix \ref{free}), and the corresponding boundary condition for the bulk modes is
\begin{align}
  \beta = \lambda_\FT \alpha \ed
  \label{bc0}
\end{align}
To compute the beta function we map the procedure from field theory to the gravity side.
We shift the renormalization scale $\mu \to \tilde\mu$ while keeping the observables (in this case the field $\phi(x,z)$) the same.
In order to keep $\phi$ the same, the modes $\a,\b$ must depend on the renormalization scale.
This leads to the relation
\begin{align}
  \phi &= \a z^\D + \b z^\D \log(\mu z) + \cdots 
  = 
  \tilde\a z^\D + \tilde\b z^\D \log(\tilde\mu z) + \cdots \ec
\end{align}
where $\tilde\a,\tilde\b$ are the modes at the shifted scale $\tilde\mu$.
We find that the relation between the original and shifted modes is
\begin{align}
  \a = \tilde\a + \tilde\b \log(\tilde\mu/\mu) \ecq \b = \tilde\b \ed
  \label{rel0}
\end{align}
We apply the boundary condition $\tilde\b = \tilde\lambda_\FT \tilde\a$ for the new modes, where $\tilde\lambda_{FT}$ is the coupling at the new scale.
Using the above relations we find that \cite{Witten:2001ua}
\begin{align}
  \tilde\lambda_\FT = 
  \frac{\lambda_\FT}{1 - \lambda_\FT \log(\tilde\mu/\mu)} \ed
\end{align}
This relation implies that the beta function for $\lambda_\FT$ is one-loop exact, and is given by
\begin{align}
  \beta_{\lambda_\FT} = \frac{d\lambda_\FT}{d\log(\mu)} = \lambda_\FT^2 \ed
\end{align}
This is a universal contribution to the double-trace beta function in the large $N$ limit, which can be seen in conformal perturbation theory \cite{Witten:2001ua}.
We will use the same method to compute other multi-trace beta functions.

\subsection{Multi-trace Beta Functions from Single-Trace Operators}
\label{suggestion}

In this paper we are interested in computing beta functions of multi-trace couplings using holography.
In particular, we are interested in beta functions that arise from single-trace interactions. 
Some examples of this were mentioned in the introduction, but we can also present a general scenario for how such beta functions can arise from single-trace operators on the field theory side, by using an argument that relates OPE coefficients and beta functions \cite{Cardy:1996xt}.\footnote{We thank Zohar Komargodski for pointing out this argument.}
Consider a set of $n$ operators $\cO_i$ with conformal dimensions $\D_i$, such that $\sum_i \D_i = d$.
Suppose there is an operator $\cO$ of dimension $d$, and that the $\cO \cO$ OPE contains a multi-trace operator, taking the form
\begin{align}
  \cO(x) \cO(0) \sim |x|^{-d} \prod_i \cO_i(0) + \cdots \ed
  \label{OPE}
\end{align}
In this scenario, turning on the marginal deformation $f \cO$ gives rise to a beta function for the multi-trace deformation $\lambda \prod \cO_i$ at order $f^2$.
Indeed, at this order in perturbation theory we bring down in the path integral
\begin{align}
  \frac{1}{2} f^2 \int \! d^dx \, \cO(x) \int \! d^dy \, \cO(y) =
  \frac{1}{2} f^2 \int \! d^dx \, d^dw \, \cO(x) \cO(x + w) \ec
\end{align}
where $w = y - x$.
When $w$ is close to 0 we can use the OPE \eqref{OPE}. 
One of the terms in the expansion is proportional to
\begin{align}
  f^2 \log(\Lambda) \int \! d^dx \, \prod_i \cO_i(x) \ec
\end{align}
where $\Lambda$ is the cutoff.
This generates a beta function for the multi-trace coupling $\lambda$.

This argument can actually be used to tell us how the multi-trace beta function should appear in the holographic dual.
Let $\phi$ be the massless scalar field that is dual to $\cO$, and let $\phi_i$ be the fields dual to $\cO_i$.
Perhaps the simplest way to generate the OPE \eqref{OPE} would be to include a bulk interaction of the form 
\begin{align} \label{bulk_coupling}
  \int \! d^{d+1}x \sqrt{g} \, \phi^2 \prod_i \phi_i \ed
\end{align}
Such an interaction contributes to the correlator $\langle \cO \cO \prod_i \cO_i \rangle$ at separated points.
If the limit in which the $\cO_i$ coincide is regular, then this interaction also leads to the desired OPE \footnote{This happens, for example, for the $\phi_1 \phi_2 \phi_3$ interaction in \cite{Freedman:1998tz}, provided that $\D_2=\D_1+\D_3$.}.
In practice, in this calculation one encounters divergences that must be subtracted, but this does not change the conclusion.

Next, we turn on the marginal deformation $f\cO$ in the field theory; this corresponds to shifting $\phi \to \phi + f$ in the bulk.
Expanding around the new vacuum, we find the bulk interaction term 
\begin{align} \label{bulk_beta}
  f^2 \int \! d^{d+1}x \sqrt{g} \, \prod_i \phi_i \ec
\end{align}
which should lead to the multi-trace beta function discussed above.
Our guess is thus that including a term of this type in the bulk action, with a coefficient $\eta$, should lead to a beta function proportional to $\eta$ for the corresponding multi-trace coupling constant.\footnote{Note that we could also generate the desired OPE \eqref{OPE} from more complicated bulk couplings that involve extra derivatives in 
\eqref{bulk_coupling}, and we would generally expect all such terms to contribute to the beta function.}

This is quite surprising, since naively one would expect a term like (\ref{bulk_beta}) to affect the correlator $\langle \prod_i \cO_i (x_i)\rangle$ at separated points, rather than having anything to do with multi-trace operators. However, we will show in the next sections that indeed terms like (\ref{bulk_beta}) give rise
(at least at linear order in their coefficient) to a multi-trace beta function. Since the same bulk coupling is usually associated with
the $n$-point function, this raises the problem of how to get $n$-point functions $\langle \prod_i \cO_i(x_i) \rangle$ that will be independent of the multi-trace
beta function (as in field theory).
We will show that
 combinations of such bulk terms, involving also terms with derivatives, contribute to the $n$-point function without affecting the beta function.
 
\section{Marginal Multi-trace Deformations}
\label{multi-trace}

In this section we consider a field theory that has a weakly coupled holographic dual and a scalar operator $\cO$ of dimension $\D = d/n$, where $n \geq 3$, such that $\cO^n$ is a marginal deformation.\footnote{
Beta functions for irrelevant multi-trace operators were considered in \cite{vanRees:2011fr,vanRees:2011ir}.}
The Euclidean bulk action for the dual field $\phi$ is taken to be
\begin{align}
  S_{\rm bulk} &= \int \! d^{d}x \int dz \, \sqrt{g}
  \left[ 
  \frac{1}{2} g^{\mu\nu} \dho_\mu \phi \dho_\nu \phi + \frac{1}{2} m^2 \phi^2
  + \frac{\theeta}{n} \phi^n
  \right] \ed
  \label{Sbulkn}
\end{align}
Guided by the discussion of section \ref{suggestion}, we have included a $\phi^n$ interaction term, whose coefficient $\eta$
could depend on various single-trace coupling constants (as in section \ref{suggestion}).
We will compute the contribution of this bulk interaction 
to the multi-trace beta function at leading order in $\eta$.

As usual in holographic computations we will encounter IR divergences in the bulk, which correspond to UV divergences in the dual field theory, so we must `renormalize' the theory \cite{Henningson:1998gx,de Haro:2000xn}
\footnote{A different approach to holographic renormalization, based on the Wilsonian
	renormalization group, was proposed in \cite{Heemskerk:2010hk,Faulkner:2010jy}. This approach discusses
	multi-trace operators more carefully, in a way that is somewhat similar to
	our computations, but we will not explicitly use it here. Results very similar to ours were found in the Wilsonian holographic
	renormalization group approach in \cite{Grozdanov:2011aa}.}.
We first regulate it by placing a cutoff at $z=\e$, and we will include a boundary action $S_{\rm ct}$ of local counter-terms to subtract divergences.
We would like to apply specific boundary conditions that correspond to multi-trace deformations, and we must therefore verify that these conditions are compatible with extremizing the action --- including its boundary piece.
We will generally have to add also non-singular boundary terms, denoted $S_\dho$, to ensure this compatibility.
Our total action will therefore be
\begin{align}
  S = S_{\rm bulk} + S_\dho + S_{\rm ct} \ed
  \label{Stot}
\end{align}

\subsection{Renormalization}

In this section we solve the bulk equations of motion to leading order in $\eta$, and introduce counter-terms to subtract divergences in the on-shell action.
The scalar equation of motion is
\begin{align}
 z^2 \phi'' + z^2 \square_x \phi + (1-d) z \phi' - m^2 \phi = 
 \theeta \phi ^{n-1}(x,z) \ec
 \label{bulkeq}
\end{align}
where $\phi' \equiv \p_z \phi(x,z)$. 
In appendix \ref{soleq} we solve this equation near the boundary, to leading order in $\eta$.
The solution (for $\Delta = d/2 - \nu$) is
\begin{align}
  \phi(x,z) &= \a(x) z^\D + \b(x) z^{d-\D} +
  \frac{\theeta}{2\nu} \a^{n-1}(x) z^{d-\D}\log(\mu z) +
  O(z^{\D+2}) + O(\eta z^{d-\Delta+2\nu}) + O(\eta^2) \ed
  \label{phisol}
\end{align}
Here $\mu$ is an arbitrary scale which we interpret as the renormalization scale.
The appearance of a logarithm hints at a loss of conformality and the generation of a beta function, and we will see that this is indeed the case.

The next step is to compute the on-shell bulk action \eqref{Sbulkn}.
After integrating by parts in the $z$ direction, and using the scalar equation of motion, we find
\begin{align}
  S_{\rm bulk}^{\rm on-shell} &=
  - \frac{1}{2} \int_{z=\epsilon} \! d^dx \sqrt{\g} \, \e \phi' \phi(x)
  + \left( \frac{2-n}{2n} \right) \theeta \int \! d^d x \int_{\epsilon} dz \sqrt{g}\,  
  \phi^n(x,z)
  \ed \label{bos}
\end{align}
Here $\g$ is the induced metric on the boundary ($\sqrt{\g} = \epsilon^{-d}$).
We now plug in the solution \eqref{phisol}, and introduce local counter-terms to subtract the divergences that appear.
The diverging part of the action as $\epsilon \to 0$ is
\begin{align}
  \int_{z=\e} \! d^dx \left[ 
  - \frac{\Delta}{2} \a^2(x) \epsilon^{-2\nu}
  - \frac{d \theeta}{4\nu} \a^n(x) \log (\mu \epsilon)
  \right] 
  +
  \frac{\nu\theeta}{d} \int_{z=\epsilon} \! d^dx \, 
  \a^n(x) \log(\mu\epsilon) 
  \ed
\end{align}
The first integral contains the divergence from the free term in \eqref{bos}, and the second integral contains the one from the interacting term.
The divergences can be subtracted by the counter-term action
\begin{align}
  S_{\rm ct} = \int_{z=\e} \! d^dx \sqrt{\g} \left[ 
  \frac{\D}{2} \phi^2(x) +
  \frac{\theeta}{n} \log(\mu_0 \epsilon) \, \phi^n(x)
  \right] 
  \label{Sct} \ed
\end{align}
The first counter-term cancels the power-law divergence, and the second one cancels the remainder of the log divergence.
We introduced an arbitrary parameter $\mu_0$ in \eqref{Sct} on dimensional grounds, which affects only the finite part of the
boundary action.

\subsection{Boundary Conditions and the Multi-trace Beta Function}

In this section we write down the complete boundary action that is compatible with the boundary conditions of a multi-trace deformation, and we compute the beta function for this deformation.
In order to turn on a multi-trace deformation of the form $W = \lambda_\FT \int d^dx \cO^n / n$ in the field theory, we need to apply the boundary condition
\begin{align}
  \b = \lambda_\FT (-2\nu\a)^{n-1}
  \label{bc}
\end{align}
for the scalar modes.
This condition must be compatible with the variation of the action.
For this purpose, let us introduce the boundary action
\begin{align}
  S_\dho = 
  \left( \frac{2\nu\l}{n} + \frac{\eta}{2\nu n} \right)
  \int_{z=\e} \! d^dx \sqrt{\g} \, \phi^n(x) \ed
  \label{Sdho}
\end{align}
(The factors in front are for later convenience.)
It is easy to see that this action does not introduce new divergences into the on-shell action.
The total action \eqref{Stot} is now given by
\begin{align}
\label{Sfull}
  S &= \int \! d^{d}x \int_\e dz \, \sqrt{g}
  \left[ 
  \frac{1}{2} g^{\mu\nu} \dho_\mu \phi \dho_\nu \phi + \frac{1}{2} m^2 \phi^2
  + \frac{\theeta}{n} \phi^n
  \right] 
  \cr &\quad
  + 
  \int_{z=\e} \! d^dx \sqrt{\g} \left[ 
  \frac{\D}{2} \phi^2(x) +
  \left( 
  \frac{\theeta}{n} \log(\mu_0 \epsilon) +
  \frac{2\nu\l}{n} + \frac{\eta}{2\nu n} \right) \phi^n(x)
  \right] 
  \ed
\end{align}
We now vary this action with respect to the scalar, and keep the variation $\delta \phi$ arbitrary.
We find that in order to extremize the action, the following equation must be satisfied on the boundary:
\begin{align}
  \D \phi - \e \phi' + \left[ \theeta \log(\mu_0 \e) + 2\nu\lambda + \frac{\eta}{2\nu}
  \right] \phi^{n-1}(x) = 0 \ed
\end{align}
Once we plug in the scalar solution \eqref{phisol}, we find that the boundary conditions are\footnote{
When writing subleading terms such as $O(\e^{2\nu})$, we are ignoring possible $\log(\e)$ factors which do not affect the discussion.}
\begin{align}
\label{phi_n_bc}
  \beta(x) = \left[ \lambda  - \frac{\theeta}{2\nu} \log(\mu/\mu_0) \right] \a^{n-1}(x)
  + O(\e^{2-2\nu}) + O(\eta \e^{2\nu}) + O(\theeta^2) \ed
\end{align}
Neglecting the subleading corrections and comparing with the desired boundary conditions \eqref{bc}, we can identify the multi-trace deformation with
\begin{align}
  \lambda_\FT = \frac{1}{(-2\nu)^{n-1}}
  \left[ \lambda   - \frac{\theeta}{2\nu} \log(\mu/\mu_0)  \right] + O(\eta^2)\ed
\label{runninglambda}
\end{align}
We see that the multi-trace coupling of the field theory corresponds to a mix of the two $\phi^n$ couplings --- in the bulk and on the boundary.

From \eqref{runninglambda} it is immediate to compute the beta function 
\begin{align}
  \b_{\lambda_{\FT}} = \frac{d\lambda_\FT}{d\log (\mu)} =
  \frac{\theeta}{(-2\nu)^n} + O(\eta^2) \ed
  \label{blcft}
\end{align}
Notice that the beta function does not depend on $\lambda_\FT$, but only on the single-trace coupling constants encoded in $\eta$.
This can be understood from large $N$ counting for $n \ge 3$: contributions of $\lambda_\FT$ to its own beta function, of order $\lambda_\FT^2$ or higher, are suppressed by factors of $1/N$.

Alternatively, we can compute the beta function by the same method we used in section \ref{sec:DoubleTrace}. 
This method uses only the bulk solution and the boundary conditions, and does not rely on the renormalization procedure or the explicit boundary action.
To do this we shift the renormalization scale $\mu$ in the bulk solution \eqref{phisol}, $\mu \to \tilde\mu$, while keeping the field $\phi(x,z)$ fixed in the bulk.
We have
\begin{align}
  \phi &= \a(x) z^{\D} + \b(x) z^{d-\D} 
  + \frac{\theeta}{2\nu} \a^{n-1}(x) z^{d-\D} \log(\mu z) 
  + O(z^{\D + 2}) + O(\eta z^{d-\D+2\nu}) + O(\theeta^2)
  \cr
  &= \tilde\a(x) z^{\D} + \tilde\b(x) z^{d-\D} 
  + \frac{\theeta}{2\nu} \tilde\a^{n-1}(x) z^{d-\D} \log(\tilde\mu z) 
  + O(z^{\D + 2}) + O(\eta z^{d-\D+2\nu}) + O(\theeta^2)
  \ec
\end{align}
where $\tilde\a,\tilde\b$ are the modes at the shifted scale $\tilde\mu$.
We find the relation
\begin{align}
 \a &= \tilde\a + O(\theeta^2) \ecq
 \b = \tilde\b +  \frac{\theeta}{2\nu} \tilde\a^{n-1} \log( \tilde\mu / \mu) 
 + O(\theeta^2) \ed
\end{align}
Next we demand that the shifted modes satisfy a boundary condition of the form $\tilde\b = \tilde\l_\FT (-2\nu\tilde\a)^{n-1}$, whre $\tilde\l_\FT$ is the effective coupling at the shifted scale $\tilde\mu$.
The couplings are then related by
\begin{align}
 \l_\FT = \tilde\l_\FT + \frac{\theeta}{(-2\nu)^n} \log(\mu/\tilde\mu) 
 + O(\eta^2) \ed
\end{align}
We find that the beta function is given by equation \eqref{blcft}, reproducing the answer we found above.

\subsection{The $n$-Point Function}
\label{nptsec}

In this section we are interested in contributions to the $n$-point correlation function of $\cO$.
A $\phi^n$ bulk interaction term will generally contribute to this correlator at leading order in the coupling (see for example \cite{Freedman:1998tz}).
When the operator dimension is $\D = d/n$, it is easy to check that this contribution has a divergence, and that this divergence is canceled by the $\phi^n$ counter-term that we introduced to renormalize the action.
In general there will be a finite contribution to the correlator after renormalization.

As we saw in the previous section, turning on a $\phi^n$ bulk interaction inevitably breaks conformal invariance when the operator dimension is $\D = d/n$ and $n \ge 3$.
Which couplings can we turn on in the gravity theory, that affect the $n$-point function of $\cO$ without breaking conformal invariance?
The boundary interaction term $\lambda \phi^n$ gives one such coupling, which affects both the $n$-point function and the multi-trace coupling $\lambda_{\FT}$ in the field theory.
As we will now see, there are also bulk couplings that have these properties.
Together with the boundary coupling $\lambda$, they can be used to independently control both the $n$-point function and the multi-trace coupling.
For the case of $n=3$, the couplings we will write down determine the 3-point function completely, while for larger $n$ they will affect a specific conformal structure in the correlator.

Let us now set $\eta = 0$ and consider the following bulk action:
\begin{align}\label{SbulkAlt}
  S_{\rm bulk} &= \int \! d^{d}x \int_\epsilon dz \, \sqrt{g}
  \left[ 
  \frac{1}{2} (\dho\phi)^2 + \frac{1}{2} m^2 \phi^2
  + \zeta \left( 
  \phi^{n-2} g^{\mu\nu} \dho_\mu \phi \dho_\nu \phi
  - \D^2 \phi^n
  \right)
  \right] \ed
\end{align}
Let us compute the contributions of $\zeta$ to the field theory $n$-point function and to the multi-trace coupling.
The variation of this action is
\begin{align}
  \delta S_{\rm bulk} &=
  \int \! d^dx \int_\e dz \sqrt{g} \, \delta \phi \left[ 
  -\square \phi + m^2 \phi - d \D \zeta \phi^{n-1}
  - (n-2) \zeta \phi^{n-3} (\dho\phi)^2
  - 2 \zeta \phi^{n-2} \square\phi
  \right]  \cr
  &\quad   \label{dSbulk}
  - \int_\dho d^dx \sqrt{\g} \, \delta\phi \left( 
  1 + 2 \zeta \phi^{n-2}
  \right) \e \phi' \ed
\end{align}
The solution of the bulk equation of motion near the boundary is 
\begin{align}
  \phi = z^\D \left[ \a(x) + O(z^2) \right] + 
  z^{d-\D} \left[ \b(x) + O(z^2) \right]
  + O(\zeta^2) + O(z^{d-\D + 2\nu}) \ed
\end{align}
The relative factor of $\D^2$ in the action \eqref{SbulkAlt} was chosen to cancel the log term that would otherwise appear at order $\zeta$.
With this choice of couplings the bulk modes are independent of the renormalization scale, and therefore the beta function of the $\cO^n$ coupling in the field theory vanishes.
In addition, the on-shell action is renormalized by a single counter-term,
\begin{align}
  S_{\rm ct} = \frac{\D}{2} \int \! d^dx \sqrt{\g} \, \phi^2 \ed
  \label{ct0}
\end{align}
In particular, we do not need to introduce counter-terms with an explicit $\zeta$ dependence, or with a scale dependence.\footnote{
Indeed, the on-shell bulk action is given by
\begin{align}
  - \int \! d^dx \sqrt{\g} \, \e \phi' \phi +
  \zeta \int \! d^d x \int_{\epsilon} dz \sqrt{g} \phi^{n-2} \left[ 
  \nu\D \phi^2 + \frac{n}{2} (\dho\phi)^2 + \phi \square \phi
  \right] \ed
\end{align}
The first integral contains a power-law divergence that is subtracted by the counter-term \eqref{ct0}, while the second integral is finite.
}

We now have a renormalized gravity theory that is dual to a conformal field theory for every value of $\zeta$.
We include two additional terms in the boundary action.
The first is the boundary interaction $\lambda \phi^n(x)$ that was discussed above.
The second is a term that is proportional to $J(x) \phi(x)$, which is necessary for compatibility with the boundary conditions (written below) that include a source for $\cO$.
The total action reads
\begin{align}
  S &= \int \! d^{d}x \int_\epsilon dz \, \sqrt{g}
  \left[ 
  \frac{1}{2} (\dho\phi)^2 + \frac{1}{2} m^2 \phi^2(x,z)
  + \zeta \left( 
  \phi^{n-2} g^{\mu\nu} \dho_\mu \phi \dho_\nu \phi
  - \D^2 \phi^n(x,z)
  \right)
  \right] 
  \cr &\quad +
  \int_{z=\e} \! d^dx \sqrt{\g} \, 
  \left[ \frac{\D}{2} \phi^2(x) 
  + 2\nu \e^{d-\D} J(x) \phi(x)
  + \frac{2\nu\l}{n} \phi^n(x) \right]
  \ed
  \label{Szeta}
\end{align}
One combination of the couplings $\lambda$ and $\zeta$ should correspond to the multi-trace deformation $\lambda_{\FT} \cO^n / n$
(which can now be turned on without any beta function).
The other combination, as we shall see, independently controls a specific conformal structure in the $n$-point function of $\cO$.

To find the combination that corresponds to a multi-trace deformation, we write down the boundary conditions that are compatible with the boundary equation that follows from the variation of the total action \eqref{Szeta}.
Using \eqref{dSbulk} we get the boundary equation
\begin{align}
  \D \phi - \e \phi' + 2\nu\e^{d-\D} J +  2\nu\lambda\phi^{n-1}
  -2 \zeta \phi^{n-2} \e\phi' = 0 \ed
\end{align}
It is compatible with imposing the boundary conditions
\begin{align}
  \beta = J + \lambda_\FT (-2\nu\a)^{n-1} \ecq
  (-2\nu)^{n-1} \lambda_\FT = 
  \lambda - \frac{\D}{\nu} \zeta \ed
  \label{comb1}
\end{align}
This gives us the mapping to the multi-trace coupling.

Next, we can compute the $n$-point function $\langle \cO(x_1) \cdots \cO(x_n) \rangle$ at leading order in these couplings.
There are two Witten diagrams, one with a $\lambda$ vertex on the boundary, the other with a $\zeta$ vertex in the bulk.
The $\lambda$ contribution is given by
\begin{align}
  - n! \cdot \frac{2\nu\lambda}{n} \int_{z=\e} d^dx \sqrt{\g}
  K_\D(\epsilon,x;x_1) \cdots K_\D(\epsilon,x;x_n) \ec
\end{align}
where $K_\D$ is the boundary to bulk propagator that is given by (see appendix \ref{free})
\begin{align}
  K_\D(z,x;x') = \frac{c z^\D}{ \left[ z^2 + (x-x')^2 \right]^\D }
  \ecq 
  c \equiv \frac{\Gamma(\D)}{\pi^{d/2} \Gamma (\D - d/2)} \ed
\end{align}
Taking $\e \to 0$, we can write the result as
\begin{align}
  -2\nu\lambda c^n (n-1)! 
  \int \frac{d^dx}{(x-x_1)^{2\D} \cdots (x-x_n)^{2\D}} 
  \ed \label{lcontrib}
\end{align}
The 2-point function of $\cO$, computed in appendix \ref{free}, is
$\langle \cO(x) \cO(0) \rangle = -2\nu c |x|^{-2\D}$.
Therefore the contribution \eqref{lcontrib} is equal to
\begin{align}
  \frac{\lambda (n-1)!}{(-2\nu)^{n-1}} 
  \int d^dx \langle \cO(x) \cO(x_1) \rangle \cdots
  \langle \cO(x) \cO(x_n) \rangle 
  \ed
\end{align}
This has the expected form of a contribution to the $n$-point function that comes from an $\cO^n$ interaction in field theory.
If we set $\zeta=0$, we see from \eqref{comb1} that this is the expected contribution due to the $\lambda_{\FT}$ multi-trace deformation.

The contribution of the $\zeta$ coupling can be computed in a similar way once we notice that the bulk coupling is secretly a surface term on-shell.
Indeed, using integration by parts and applying the equations of motion, $\square \phi = m^2 \phi + O(\zeta)$, it is easy to check that
\begin{align}
  \int d^d x \int_{\epsilon} dz \sqrt{g} \left[ 
  \phi^{n-2} (\dho\phi)^2 - \D^2 \phi^n
  \right] =
  \frac{\D}{\D-d} \int_{z=\e} d^dx \sqrt{\g} \e \phi' \phi^{n-1}
  + O(\zeta) \ed
\end{align}
Using the same method as for the $\lambda$ contribution, we find that the $n$-point function at leading order is
\begin{align}
  -2\nu \left[  
  \lambda - \frac{\D}{\nu} \left( \frac{d}{d + 2\nu} \right) \zeta 
  \right]
  c^n (n-1)! \int 
  \frac{d^dx}{(x-x_1)^{2\D} \cdots (x-x_n)^{2\D}} \ed
\end{align}
We thus find that the combination of couplings that controls that particular conformal structure in the $n$-point function can be written as
\begin{align}
  \lambda - \frac{\D}{\nu} \left( \frac{d}{d + 2\nu} \right) \zeta \ed
  \label{comb2}
\end{align}
This combination and the combination \eqref{comb1} that controls the multi-trace deformation are inequivalent.
We conclude that by turning on both $\lambda$ and $\zeta$ we can independently control both the multi-trace deformation and the $n$-point function in the field theory.
For $n=3$ the correlator has a single conformal structure, so we can control it completely.
For $n>3$, the couplings we wrote down only control the conformal structure that appears in \eqref{lcontrib}.
These statements still hold true when we turn on $\theeta$ and break conformal invariance.

Note that in a specific theory, the coefficients of the two terms proportional to $\zeta$ in \eqref{Szeta} will be two
independent functions of the single-trace couplings. Our discussion in this section implies that, restricting the
bulk action to include these specific terms, the beta function for the multi-trace deformation will vanish only when
the coefficients of these two terms are equal to each other. Whenever this does not happen for any value of the
exactly marginal single-trace couplings (if any), conformal invariance will necessarily be broken.

\section{Marginal Double-trace Deformations}
\label{doubletrace1}

In this section we consider marginal double-trace deformations, and write down bulk interaction terms that turn on beta functions for these deformations.

\subsection{Marginal Double-trace Deformations with a Single Operator}

The first case we discuss involves an operator $\cO$ of dimension $\D = d/2$.
It is dual to a bulk scalar $\phi$ with mass squared given by $m^2 = -d^2/4$.
Following the strategy laid out in section \ref{suggestion}, we turn on a bulk term proportional to $\phi^2$ (this is the double-trace analog of the $\phi^n$ interaction we considered in section \ref{multi-trace}), and compute the beta function of the $\cO^2$ coupling.
The bulk action is
\begin{align}
  S_{\rm bulk} &= \frac{1}{2} \int \! d^{d}x \int_\epsilon dz \, \sqrt{g}
  \left[ 
  (\dho\phi)^2 + m^2 \phi^2 + \eta \phi^2
  \right] \ed
  \label{ddSbulk}
\end{align}
The $\eta$ deformation shifts the mass, so we must choose $\eta \ge 0$ in order not to violate the Breitenlohner-Freedman bound \cite{Breitenlohner:1982jf}.
The reader may wonder how such a shift can break conformal invariance in a non-trivial way; as we will discuss below, there are indeed boundary
conditions on $\phi$ for which conformal invariance is preserved, but they are not the ones we get by starting from the boundary
conditions corresponding to $\Delta=d/2$. For now let us treat $\eta$ as a perturbation and solve to leading order.
Solving the bulk equation of motion near the boundary, we find
\begin{align}
  \phi &= \alpha(x) z^{d/2} + \beta(x) z^{d/2} \log(\mu z)
  + \frac{\eta}{2} \alpha(x) z^{d/2} \log^2(\mu z)
  + \frac{\eta}{6} \beta(x) z^{d/2} \log^3(\mu z) 
  \cr &\quad
  + O(z^{d/2+2}) + O(\eta^2) \ed
  \label{ddphi}
\end{align}
The field theory deformation $W = -\lambda_\FT \int \! d^dx \, \cO^2 / 4$ corresponds to the boundary conditions\footnote{
The relation between the expectation value and the mode $\alpha$ in this case is $\langle \cO \rangle = -2\alpha$; see appendix \ref{free}.}
\begin{align}
  \beta = \lambda_\FT \alpha \ed
  \label{ddbc}
\end{align}
As before, we compute the beta function by varying $\mu$ while keeping $\phi$ constant, and the result is
\begin{align}
  \beta_{\lambda_\FT} = \lambda_\FT^2 - \eta + O(\eta^2) \ed \label{ddbeta}
\end{align}
This contains the usual universal $\lambda_\FT^2$ term that we discussed in section \ref{sec:DoubleTrace}. However, now
the beta function does not vanish if we set $\lambda_\FT=0$, so turning on the bulk coupling $\eta$ (which we interpret as some
function of the single-trace couplings of the field theory) breaks conformal invariance as expected.
On the other hand, in this case we also have two fixed points, which are at $\lambda_\FT = \pm \sqrt{\eta}$ to leading order.

Alternatively, we can solve exactly in $\eta$.
The bulk field is given by
\begin{align}
  \phi = z^{\Delta} \left[ \tilde\a(x) + O(z^2) \right]
  + z^{d - \Delta} \left[ \tilde\b(x) + O(z^2)  \right]\ec
\end{align}
where $\Delta = d/2 - \nu$ and $\nu = \sqrt{\eta}$.
The new modes are related to the previous ones by 
\begin{align}
  \a = \mu^{-\nu} \tilde\b + \mu^{\nu} \tilde\a \ecq
  \b = \nu
  \left( \mu^{-\nu} \tilde\b - \mu^{\nu} \tilde\a \right) \ed
  \label{modesrel}
\end{align}
In terms of these new modes, the boundary condition \eqref{ddbc} is
\begin{align} \label{relation}
  \tilde\beta = \mu^{2\nu} g \tilde\alpha \ecq 
  g \equiv \frac{\nu + \lambda_\FT}{\nu - \lambda_\FT} \ed
\end{align}
We now have an operator $\tilde{\cO}$ of dimension $\Delta < d/2$, and $\mu^{2\nu} g$ controls the relevant deformation $\tilde{\cO}^2$ of dimension
$2 \Delta = d - 2 \nu$.
The running of the dimensionless coupling $g$ is determined by this dimension, and its beta function is given by
\begin{align}
  \beta_g = -2\nu g \ed
\end{align}
This agrees with \eqref{ddbeta} to leading order in $\eta$.

The two fixed points, at $\lambda_{\FT} = \pm \nu$, correspond using \eqref{relation} to $g=0$ or $g=\infty$.
These give the boundary conditions $\tilde\beta=0$ and $\tilde\alpha=0$, respectively, which are the two quantizations where we don't turn on a double-trace deformation.
These correspond to two conformal field theories. As usual one can flow from the $\tilde\beta=0$ CFT to the $\tilde\alpha=0$ CFT by turning on the ${\tilde \cO}^2$ deformation in that theory. But we see that the boundary condition that we get by starting from the theory
with $\Delta=d/2$ and no double-trace deformation, $\lambda_{\FT}=0$ or $g=1$, is far from these fixed points, and as we saw it does lead
to a breaking of conformal invariance by a beta function for the double-trace operator. The end-point of the corresponding renormalization group (RG) flow
is precisely the CFT that we get by quantizing the same bulk action with the boundary condition $\tilde\alpha=0$, in which the scalar
field corresponds to an operator of dimension $\Delta=d/2+\sqrt{\eta}$. This has an alternative description as the theory with
$\Delta=d/2$ that we started from (which was not conformal), with a finite double-trace deformation (and possibly other effects
related to the bulk interaction proportional to $\eta$).

\subsection{Marginal Double-trace Deformations with Two Operators}
\label{doubletrace2}

In this subsection we consider marginal double-trace deformations involving two different operators $\cO_1$ and $\cO_2$ with dimensions $d - \D$ and $\D$, respectively, such that $(d/2) -1<\D < d/2$, 
and write down bulk interaction terms that turn on beta functions for these deformations.
As usual, we denote $\D = \frac{d}{2} - \nu$.
The operators are dual to bulk fields $\phi_1,\phi_2$, both with the same mass squared, $m^2 = \D (\D-d)$.
As before, the action will have three pieces,
\begin{align}
  S = S_{\rm bulk} + S_{\rm ct} + S_\dho \ec
\end{align}
where $S_{\rm ct}$ contains the counter-terms, and $S_\dho$ are regular boundary terms.
The bulk action for this theory is taken to be
\begin{align}
  S_{\rm bulk} &= \frac{1}{2} \int d^dx \int_\epsilon dz \sqrt{g} \left[ 
  (\dho \phi_1)^2 + (\dho \phi_2)^2 
  + m^2 \phi_1^2 + m^2 \phi_2^2
  + 2 \eta \phi_1 \phi_2
  \right] \ed
  \label{Sbulk1}
\end{align}
As we will show, the term $\phi_1 \phi_2$ will generate a beta function for the double-trace deformation $\cO_1 \cO_2$ at order $\eta$, even when the double-trace coupling vanishes.
The reasoning for choosing this term again follows from the discussion in section \ref{suggestion}.

As before, the same action can also be quantized as a conformal theory in which the dimensions of the operators
are shifted at order $\eta$.
Indeed, after diagonalizing the mass matrix we find that the decoupled fields $\psi_\pm = \phi_1 \pm \phi_2$ can correspond to operators with dimensions
\begin{align}
  \D_\pm = \frac{d}{2} \pm \nu + \frac{\eta}{2\nu} + O(\eta^2)
  \label{Dpm}
\end{align}
in a theory with no double-trace couplings.
However, in our case we will choose different boundary conditions, corresponding to the original scaling dimensions
with a small beta function for the double-trace deformation; as before, the conformal field theory with both dimensions equal to  $\Delta_+$ in \eqref{Dpm}
will arise at the end of the RG flow.

\subsubsection{Renormalization}

We start by solving the bulk equations of motion, and renormalizing the theory.
Renormalization is not strictly necessary for computing the beta function, and we include this discussion for completeness.
The variation of the bulk action is
\begin{align}
  \delta S_{\rm bulk} &= 
  \int d^dx \int_\epsilon dz \sqrt{g} \left[
  \delta \phi_1 (-\square \phi_1 + m^2 \phi_1 + \eta \phi_2) +
  \delta \phi_2 (-\square \phi_2 + m^2 \phi_2 + \eta \phi_1)
  \right] \cr
  &\quad - \int d^dx \sqrt{\g} \left( 
  \delta\phi_1 \epsilon \phi_1' + \delta\phi_2 \epsilon \phi_2'
  \right) \ed
\end{align}
The solution of the bulk equations of motion near the boundary (see appendix \ref{soleq}) is
\begin{align}
  \phi_1 &= \a_1(x) z^\D + \b_1(x) z^{d-\D}
  + \frac{\eta}{2\nu} \left[ \b_2(x) z^{d-\D} - \a_2(x) z^\D \right]
  \log(\mu z) 
  + O(z^{\D+2}) + O(\eta^2) \ec\cr
  \phi_2 &= \a_2(x) z^\D + \b_2(x) z^{d-\D}
  + \frac{\eta}{2\nu} \left[ \b_1(x) z^{d-\D} - \a_1(x) z^\D \right]
  \log(\mu z) 
  + O(z^{\D+2}) + O(\eta^2) \ed
\end{align}
The bulk on-shell action is given by
\begin{align}
  S_{\rm bulk}^{\rm on-shell} &=
  - \frac{1}{2} \int_{z=\e} d^dx \sqrt{\g} \e (\phi_1 \phi_1'
  + \phi_2 \phi_2') \cr
  &= -\frac{1}{2} \int_{z=\e} d^dx \e^{-2\nu} \left[ 
  \D (\a_1^2 + \a_2^2) - \frac{2\eta}{\nu} \D \a_1 \a_2 \log(\mu\e)
  - \frac{\eta}{\nu} \a_1 \a_2
  \right] + ({\rm finite}) + O(\eta^2) \ed
  \cr
\end{align}
The following counter-term action cancels the divergences:
\begin{align}
  S_{\rm ct} = \frac{1}{2} \int d^dx \sqrt{\g} \left[ 
  \D (\phi_1^2 + \phi_2^2) - \frac{\eta}{\nu} \phi_1 \phi_2
  \right] \ed
\end{align}

\subsubsection{Boundary Conditions and Beta Function}

Let us write down the boundary conditions that correspond to a double-trace deformation, and compute the beta function for the double-trace coupling.

On the field theory side we introduce the term $W = -\lambda_{\FT} \cO_1 \cO_2$.
Using \eqref{manyBC}, this corresponds to the boundary conditions 
\begin{align}
  \alpha_1 = - 2\nu \lambda_\FT \alpha_2 + O(\eta) \ecq
  \beta_2 = 2\nu \lambda_\FT \beta_1 + O(\eta) \ed
  \label{dtbc}
\end{align}
The extra minus sign is due to the fact that one of the dimensions is greater than $d/2$.
The $O(\eta)$ terms correspond to possible corrections coming from the extra term in the bulk action.
They will not affect the result for the beta function at leading order in $\eta$ so we can ignore them.

To compute the beta function we again shift the renormalization scale, taking $\mu \to \tilde\mu$ while keeping $\phi$ fixed and keeping track of the bulk modes.
The $\phi_1$ field for the two choices can be written as
\begin{align}
  \left.\phi_1\right|_{\mu} &= \a_1 z^\D + \b_1 z^{d-\D}
  - \frac{\eta}{2\nu} (\a_2 z^\D - \b_2 z^{d-\D}) \log(\mu z) 
  + O(z^{\D+2}) + O(\eta^2) \ec
  \cr
  \left.\phi_1\right|_{\tilde\mu} &= \tilde\a_1 z^\D + \tilde\b_1 z^{d-\D}
  - \frac{\eta}{2\nu} (\tilde\a_2 z^\D - \tilde\b_2 z^{d-\D}) \log(\tilde\mu z) 
  + O(z^{\D+2}) + O(\eta^2)
  \ec
\end{align}
and similarly for $\phi_2$.
Demanding that $\left.\phi_i\right|_{\mu} = \left.\phi_i\right|_{\tilde\mu}$, we find the relations
\begin{align}
\label{shifts}
  \tilde\a_1 &= \a_1 + \frac{\eta}{2\nu} \log(\tilde\mu/\mu) \a_2 + O(\eta^2) \ec &
  \tilde\a_2 &= \a_2 + \frac{\eta}{2\nu} \log(\tilde\mu/\mu) \a_1 + O(\eta^2) \ec \cr
  \tilde\b_1 &= \b_1 - \frac{\eta}{2\nu} \log(\tilde\mu/\mu) \b_2 + O(\eta^2) \ec &
  \tilde\b_2 &= \b_2 - \frac{\eta}{2\nu} \log(\tilde\mu/\mu) \b_1 + O(\eta^2) \ed
\end{align}
Let us apply boundary conditions of the form \eqref{dtbc} also at $\tilde\mu$, denoting the effective coupling at this scale by $\tilde\l_\FT$.
Using \eqref{shifts}, we see that the couplings at $\mu$ and $\tilde\mu$ are related by
\begin{align}
  \tilde\lambda_\FT - \lambda_\FT = 
  \frac{\eta}{(2\nu)^2} \log(\tilde\mu/\mu) 
  \left[ (2\nu \lambda_\FT)^2 - 1 \right]
  + O(\eta^2) \ed
\end{align}
Here we used the fact the $\tilde\lambda_\FT = \lambda_\FT + O(\eta)$.
We find that the beta function is
\begin{align}
  \beta_{\lambda_\FT} = \frac{d\lambda_\FT}{d\log(\mu)} =
  \eta \left[ \lambda_\FT^2 - \frac{1}{(2\nu)^2} \right] + O(\eta^2) \ed
\end{align}
For $\lambda_\FT=0$ the situation is analogous to what we found in the higher-trace cases, discussed in section \ref{multi-trace}.
The beta function in this case is similar to the one obtained there --- see eq. \eqref{blcft} --- and we see that the bulk $\phi_1 \phi_2$ term breaks conformality.

As in the previous double-trace case, here we also find a term proportional to $\lambda_\FT^2$, whose
appearance in the double-trace case can be understood from large $N$ counting: double-trace contributions to double-trace beta functions are not suppressed at large $N$.
Notice that there are two fixed points, at $\lambda_\FT^2 = (2\nu)^{-2}$.
Again, we interpret these fixed points as corresponding to the `standard' conformal quantizations of the action \eqref{Sbulk1}.

\section {Beta Function from the stress tensor}
\label{stress}
We will now calculate the multi-trace beta function using a different approach, to back up the previous results. We will use the relation between the beta function and the trace of the stress tensor in the field theory. The stress tensor can be calculated from the gravity action by the procedure outlined in \cite{Balasubramanian:1999re,Myers:1999psa,Skenderis:2000in}, which we will review shortly, and the beta function can then be computed from a Ward identity. Our results confirm the previous result \eqref{blcft}.

Beta functions are related to the traced stress tensor by the Ward identity
\begin{align}
\langle T_i^i(x) \rangle_J = \sum_{\lambda_\FT} \beta_{\l_\FT} \langle \cO_{\l_\FT}(x) \rangle_J \ec
\label{Ward}
\end{align}
where the subscript $J$ denotes the presence of sources, and the sum is over interaction terms $\lambda_{\FT} \cO_{\lambda_\FT}$ whose beta functions are $\beta_{\lambda_\FT}$.
Here we focus on terms of the form $\lambda_{\FT} \cO^n / n$, where $\cO^n$ is a multi-trace operator.
We then expect a holographic calculation of the traced stress tensor to produce
\begin{align}
  \langle T_i^i \rangle = 
  \frac{\beta_{\l_\FT}}{n} \langle \cO \rangle^n =
  \frac{\beta_{\l_\FT}}{n} (-2 \nu \alpha)^n \ec
  \label{StressBeta}
\end{align}
where $\alpha(x)$ is the fluctuating mode of the scalar field dual to $\cO$.
Here we used large $N$ factorization and the relation $\langle \cO(x) \rangle =- 2 \nu \alpha(x)$.

We start with the matter action \eqref{Sfull} for $n \geq 3$ and add to it a source term, so our action reads
\begin{align}
\label{Sfull2}
  S_{\rm matter} &= \int \! d^{d}x \int_\e dz \, \sqrt{g}
  \left[ 
  \frac{1}{2} g^{\mu\nu} \dho_\mu \phi \dho_\nu \phi + \frac{1}{2} m^2 \phi^2
  + \frac{\theeta}{n} \phi^n
  \right] +
  \cr &\quad
  \int_{z=\e} \! d^dx \sqrt{\gamma}  \bigg[
  \frac{\D}{2} \phi^2(x) +
  2\nu \e^{(d-\D)/2} J \phi \, +
  \cr &\qquad \qquad \qquad \qquad \qquad
  \left( 
  \frac{\theeta}{n} \log(\mu \epsilon) -
  (-2\nu)^{n} \frac{\l_\FT}{n} +
  \frac{\eta}{2\nu n} 
  \right) \phi^n(x) 
  \bigg]
  \ed
\end{align}
Here we used \eqref{runninglambda}.
To calculate the field theory stress tensor we will also need to consider the contribution from the pure-gravity action,
\begin{align}
  S_{\rm gravity} &= \frac{1}{16\pi G_N} 
  \int \! d^{d}x \int_\e dz \sqrt{g} \left( R[g] + 2 \Lambda \right)
  - \frac{1}{16\pi G_N} \int_{z=\e} \! d^dx \sqrt{\gamma} \, 2 K 
   \ed
  \label{Sgrav}
\end{align}
Here $\gamma_{ij} = \left.g_{ij}\right|_{z=\e}$ is the induced metric on the boundary,
\begin{align}
  K_{ij} \equiv \frac{1}{2} (\nabla_{i} n_{j} + \nabla_{j} n_{i})
\end{align}
where $n_i$ is the unit vector normal to the boundary, and $K = \gamma^{ij} K_{ij} = \nabla^i n_i$ is the extrinsic curvature (the covariant derivative is defined with respect to the full metric $g$).
For consistency with \cite{de Haro:2000xn}, we we will work with the coordinate $\rho = z^2$, for which the metric can be written as\footnote{
In our convention the metric components $g_{\mu\nu},h_{ij}$ are dimensionless, the coordinates $x^\mu$ have length dimension, $\rho$ has length-squared dimension, and the $AdS$ space has unit radius.
}
\begin{align}
  ds^2 &= g_{\mu\nu} dx^\mu dx^\nu = 
    \frac{d \rho^2}{4\rho^2} + \frac{1}{\rho} h_{ij}(x,\rho) dx^i dx^j
  \ed 
\end{align}

The holographic dictionary tells us that the stress tensor of the dual theory is given by
\begin{align}
  \langle T_{ij} \rangle &=
  \frac{2}{\sqrt{h_0}} \frac{\delta S_{\rm ren}}{\delta h_{0}^{ij}} \ec
\end{align}
where $S_{\rm ren}$ is the renormalized action of the gravity theory \eqref{Sfull2} (obtained after subtracting divergences), and $h_0$ is the field theory metric which is the boundary value of $h_{ij}(x,\rho)$, namely $h_0^{ij}=h^{ij}(x,\e^2)$.
In the presence of a boundary at $\rho=\epsilon^2$, this can be written as
\begin{align}
  \langle T_{ij} \rangle &=
  \lim_{\epsilon \to 0} \frac{2}{\sqrt{h(x,\epsilon^2)}} 
  \frac{\delta S_{\rm ren}}{\delta h^{ij}(x,\epsilon^2)} 
  = \lim_{\epsilon \to 0} \left[ \frac{1}{\epsilon^{d-2}} 
  T^{\BY}_{ij}[\gamma] \right]
  \ed
\end{align}
Here, 
$\gamma_{ij} = \epsilon^{-2} h_{ij}$ is the induced metric at $\rho=\epsilon^2$, and
\begin{align}
  T^{\BY}_{ij}[\gamma] = \frac{2}{\sqrt{\gamma}} 
  \frac{\delta S_{\rm ren}}{\delta \gamma^{ij}}
  \label{TBY}
\end{align}
is the stress tensor of the bulk theory with a cutoff at $\rho=\epsilon^2$, also known as the Brown-York stress tensor.
After taking the trace we get
\begin{align}
\label{BYLimit}
  \langle T_i^i\rangle =
  h_0^{ij} \langle T_{ij} \rangle =
  \lim_{\e \rightarrow 0}\left[ \frac{1}{\e^{d}} \gamma^{ij} T^{\BY}_{ij} \left[\gamma \right] \right] \ed
\end{align}

To compute the stress tensor of the boundary theory we will evaluate it in terms of the scalar modes. To do that we use the scalar solution \eqref{phisol}, together with the leading order back-reaction on the metric. It will be enough to solve for traces of the metric modes. 

We will now rederive equation (5.17) of \cite{de Haro:2000xn}, which relates the metric and scalar modes.
The Einstein equation is 
\begin{align}
 \label{eq:GravityEom1}
 R_{\mu \nu} - \frac{1}{2}g_{\mu \nu} \left( R + \Lambda \right) = -8 \pi G_N T_{\mu \nu}^{\bulk} \ec
\end{align}
where $T_{\mu \nu}^{\bulk}$ is the bulk stress tensor of the matter fields, defined by
\begin{align}
 T_{\mu \nu}^{\bulk} = \frac{2}{\sqrt{g}} \frac{\delta S_\bulk}{\delta g^{\mu\nu}} \ed
\end{align}
It is useful to write down the following combination of the $\r \r$ component and the traced-$i j$ component of equation \eqref{eq:GravityEom1},
\begin{align}
 4 \left(1-d\right)\r R_{\r \r} - \frac{\Lambda}{\r} 
  = -8\pi G_N \left[h^{i j}T_{i j}^{\bulk} 
  + 4 \left(2-d\right) \r T_{\r \r}^{\bulk} \right] \ed
\end{align}
In terms of the metric modes we find the equation\footnote{For consistency with \cite{de Haro:2000xn} we work with the following conventions $R_{ijk}^{\phantom{ijk} l}=\p_{[i} \Gamma_{j] k}^l + \Gamma_{p [i}^l \Gamma_{j] k}^p$, $R_{ij}=R_{ikj}^{\phantom{ikj} k}$, where square brackets denote anti-symmetrization with respect to indices.} 
\begin{align}
\label{eq:SkenderisEMT}
\trace \left[ h^{-1} h_{,\r\r} - \frac{1}{2}h^{-1}h\dr h^{-1} h\dr \right]
  = -\frac{4 \pi G_N}{\r \left(1-d\right)} \left[h^{i j}T_{i j}^{\bulk} + 4 \left(2-d\right) \r T_{\r \r}^{\bulk} \right] \ec
\end{align}
where $h_{ij,\r} = \dho_\rho h_{ij}$.
On the left-hand side we use matrix notation for the metric.

In the absence of matter fields, the solution to this equation is given by the Fefferman-Graham expansion of $h_{ij}$,
\begin{align}
\label{eq:FGmetric}
 h(x,\rho) &= h_0(x) + \rho h_2(x) + \cdots + \rho^{d/2} h_d(x) + k_d(x) \rho^{d/2} \log (\mu^2 \rho) + \cdots \ec
\end{align}
where the logarithm only appears for even $d$ and we omitted the indices for simplicity. However, in the presence of the scalar, the various powers of $\r$ on the right-hand side of \eqref{eq:SkenderisEMT} change this expansion (see, for example, \cite{Hung:2011ta}). 
For the non-minimally coupled scalar \eqref{Sfull2}, the equation can be written as
\begin{align}
\label{Einsteinn}
\rho^2
 \trace \left[h^{-1} h_{,\r\r} +\frac{1}{2} h^{-1}\dr h\dr\right] = 
 -16 \pi G_N \left[ (\r \phi')^2 -\frac{\D^2}{4 (d-1)}\phi^2 + \frac{\eta}{2 n (d-1)} \phi^n\right] \ed
\end{align}
The bulk scalar solution is given by \eqref{phisol}. 

To solve for the metric in the presence of the scalar, we start by writing the solution as a sum of the pure-gravity piece \cite{de Haro:2000xn} and a scalar back-reaction piece,
\begin{align}
 h=h^\pure+h^\scalar\ed
\end{align}
To solve equation \eqref{Einsteinn} we will match powers of $\r$ on both sides.
As discussed in \cite{Hung:2011ta}, the scalar changes the mode expansion of the metric.
Following \cite{Hung:2011ta} we assume that 
\begin{align}
p+q \frac{\D}{2} \neq \frac{d}{2}
\label{assump}
\end{align}
for all integer $q \ge 2$ and integer $p$.\footnote{
In \cite{Hung:2011ta} it is also assumed that $\Delta$ is rational, which is always true in the cases we discuss in this section.
}
Under this assumption our mode expansion takes a form similar to (2.23) in \cite{Hung:2011ta},
\begin{align}
\label{eq:MetricExpansionLogs}
 h_{i j}(x,\rho) &= h_0(x) + \rho h_2(x) + \cdots + \rho^{d/2} h_d(x) 
 \cr &\quad 
 + \rho^{d/2} \log (\mu^2 \rho) k_d(x) + \r^{\D} \tilde{k}_{2\D}(x) + \cdots \ed
\end{align}

The field theory stress tensor for the pure gravity case was calculated in \cite{de Haro:2000xn} in terms of the metric modes, with the result
\begin{align}
\label{eq:TracePureGravity}
 \langle T_i^i \rangle = \frac{1-d}{8 \pi G_N} \trace \left[ \frac{d}{2}h_d + k_d  \right] + X\left[h_{n<d} \right] \ed
\end{align}
The terms in $X$ are related to anomalies, and $h_d,k_d$ are the only terms that survive the $\e \to 0$ limit after subtracting the divergences. 
To compute the trace of the stress tensor we need to solve \eqref{Einsteinn} only for the modes $h_d$, $k_d$, and $\tilde{k}_{2\D}$, where the latter is necessary for canceling a divergence which occurs when taking the $\e \to 0$ limit of the Brown-York tensor.
Under the assumption \eqref{assump}, equation \eqref{Einsteinn} reduces to 
\begin{align}
 \r^2 \trace \left[ h^\scalar_{,\r\r} \right] = 
 -16 \pi G_N\left[ (\r \phi')^2 -\frac{\D^2}{4 (d-1)}\phi^2 + \frac{\eta}{2 n (d-1)} \phi^n\right]
 \label{treq}
\end{align}
for the powers of $\rho$ that multiply the three modes we are interested in.
After substituting the scalar solution \eqref{phisol}, we find the solution
\begin{align}
 h_d^\scalar&=\frac{8 \pi G_N}{1-d} \left[ 4 \left(\frac{n-1}{n^2}\right)  d  \a\b -4 \eta \frac{\a^n}{d n} \right]+ O\left( \eta^2 \right)\ec \nn \\
 k_d^\scalar & =\frac{8 \pi G_N}{d-1} \left[ 2 \eta \left( \frac{n-1}{n-2} \right) \frac{\a^n}{n} \right]+O\left( \eta^2 \right)\ec \\ 
 \tilde{k}_{2\D}^\scalar&=\frac{8 \pi G_N}{1-d} \left[ \frac{d}{2} \a^2 \right]+O\left( \eta^2 \right) \ed \nn 
\end{align}

The Brown-York tensor \eqref{TBY} for the scalar with action \eqref{Sfull2}  is
\begin{align}
\label{TBYn}
  T^{\BY}_{i j}[\gamma] &= \frac{1}{8 \pi G_N} 
  (  \gamma_{i j} K - K_{i j}) + 
  \cr &\quad
  \gamma_{ij}\left[ \frac{\D}{2}\phi^2 + \left( 
  \frac{\theeta}{2\nu}
  + \theeta \log(\mu\e) - (-2 \nu)^n \l_\FT \right)
  \frac{\phi^n}{n}+2 \nu \e^{d-\D} J \phi 
  \right]  \ed
\end{align}
Let us first compute $K$ in terms of the modes of $h_{ij}$ (we will only need the trace of $K_{ij}$ to compute the trace of the stress tensor).
The normal vector is $n^\rho = 2\rho$ with all other components vanishing (it is normalized using $g_{ij}$).
We find that\footnote{
We use the Christoffel symbol 
$\Gamma_{i\rho}^j = \frac{1}{2} h^{jk} h_{ki,\r} - \frac{1}{2\rho} \delta^j_i$ of the metric $g$.
}
\begin{align}
  K = \nabla^i n_i = \dho^i n_i + \Gamma^i_{i\mu} n^\mu
  = \Gamma^i_{i\rho} n^\rho
  = \rho \trace( h^{-1} h\dr ) - d \ed
\end{align}
As in the pure-gravity case, the $\r^{d/2}$ term in $K$ is 
\begin{align}
 \r^{d/2}\left[ \frac{d}{2}h_d + k_d\right]\ec
\end{align}
and the diverging $\r^{d/2}\log (\r)$ term cancels with the contributions from the scalar (which are in the second line of \eqref{TBYn}). 
After substituting in the modes of the metric and the scalar solution we apply the multi-trace boundary condition 
\begin{align}
 \b = \l_\FT (-2\nu \a)^{n-1} +J \ed
\end{align}
To get the field theory stress tensor we need to take the limit $\e \to 0$. 
The counter-terms for the pure gravity contribution were already calculated in equation (3.3) of \cite{de Haro:2000xn}, and we encounter no new additional divergences when taking the limit. 
Thus we find that the contribution to the field theory stress tensor from the back-reacted modes is
\begin{align}
 \langle T_i^i \rangle_J &=  J \left( d- \D\right) 2\nu \a + \frac{\eta}{n} \a^n +X\left[ h_{n<d}\right]+O\left(\eta^2\right) \ec
\end{align}
where again $X$ is a function of the pure-gravity metric modes $h_{n<d}$ as found in \cite{de Haro:2000xn}.
From here we can read off the beta function using \eqref{StressBeta}, and we find
\begin{align}
 \beta_{\l_\FT} = \frac{\eta}{(-2\nu)^n} \ed
\end{align}
This is in accord with the previous result \eqref{blcft}.

For the case of $n=2$ (double-trace deformations) the computation is similar and reproduces \eqref{ddbeta}.

\section*{Acknowledgements}
We would like to thank Lorenzo Di Pietro, Xi Dong, Ethan Dyer, Zohar Komargodski,  Stephen Shenker, and Eva Silverstein for useful discussions. This work was supported in part by an Israel Science Foundation center for excellence grant, by the Minerva foundation with funding from the Federal German Ministry for Education and Research, by the I-CORE program of the Planning and Budgeting Committee and the Israel Science Foundation (grant number 1937/12), by a Henri Gutwirth award from the Henri Gutwirth Fund for the Promotion of Research, by the ISF within the ISF-UGC joint research program framework (grant no. 1200/14), and by a grant from the John Templeton Foundation. The opinions expressed in this publication are those of the authors and do not necessarily reflect the views of the John Templeton Foundation. OA is the Samuel Sebba Professorial Chair of Pure and Applied Physics.

\appendix

\section{Free Bulk Scalars and Holography}
\label{free}

In this appendix we review some known results for bulk scalar fields that are dual to operators with dimension $(d/2) - 1 < \Delta < (d/2) + 1$, with no multi-trace deformations.
These include the boundary-to-bulk propagator, the renormalized on-shell action, and the precise relation between the operator expectation value and the fluctuating mode of the scalar \cite{Klebanov:1999tb,Witten:1998qj,Hartman:2006dy}.

Consider a scalar operator $\cO$ of dimension $\D = (d/2) - \nu$, which is dual to a scalar field $\phi$ in the Poincar\'e patch of $AdS_{d+1}$.
The action is
\begin{align}
  S = \frac{1}{2} \int d^dx \int_0^\infty dz \sqrt{g} \left[ 
  (\dho \phi)^2 + m^2 \phi^2(x,z)
  \right] + S_{\mathrm{ct}} \ecq
  m^2 = \Delta (\Delta - d) \ec
  \label{Sfree}
\end{align}
where the metric was defined in \eqref{metric}.
The action $S_{\mathrm{ct}}$ will contain boundary terms that are required for renormalization and for compatibility with the boundary conditions.

\subsection{The Case $\Delta \ne d/2$}

When $\Delta \ne d/2$ the field has the near-boundary expansion
\begin{align}
  \phi(x,z) = z^\D \left[ \a(x) + O(z^2) \right] 
  + z^{d-\D} \left[ \b(x) + O(z^2) \right] \ec
  \label{phifree}
\end{align}
and we identify $\beta$ with the source $J$ of the operator.
Namely, the holographic dictionary for the generating function of $\cO$ correlators is
\begin{align}
  Z[J] =
  \left\langle \exp \int \! d^dx \, J(x) \cO(x) \right\rangle =
  \exp \left( -S_{\mathrm{on-shell}}(\phi) \right)
  \ec \label{dict}
\end{align}
where the on-shell action is evaluated with the boundary condition $\beta(x) = J(x)$.\footnote{
This is consistent with our convention for the multi-trace boundary conditions, equations \eqref{SW} and \eqref{bcdef}.
}
In what follows we will compute the on-shell action, and derive the relation
\begin{align}
  \langle \cO(x) \rangle = -2\nu \alpha(x)
\end{align}
between the expectation value of the operator and the fluctuating mode of the scalar.

\subsubsection{Boundary-to-Bulk Propagator}

We begin by computing the boundary-to-bulk propagator, $K_\D(x,z;x')$. 
It is defined by the condition that
\begin{align}
  \phi(x,z) = \int d^dx' K_\D(x,z;x') \beta(x')
\end{align}
is a solution of the bulk equation of motion $\square \phi - m^2 \phi = 0$, such that the coefficient of the $z^{d-\D}$ mode is $\beta(x)$.
By setting $\beta(x) = \delta(x-x_0)$ we see that $K_\D(x,z;x_0)$ is itself a solution of the equation of motion, which includes the specific mode $\delta(x-x_0) z^{d-\D}$.

Due to translation invariance we have $K_\D(x,z;x') = K_\D(x-x',z)$.
It is convenient to first compute the momentum-space propagator,
\begin{align}
  K_\D(k,z) = \int d^dx e^{-ik\cdot x} K_\Delta(x,z) \ed
\end{align}
It satisfies the bulk equation of motion
\begin{align}
  z^2 \dho_z^2 K_\D + (1-d) z\dho_z K_\D 
  - (m^2 + z^2 k^2) K_\D(k,z) = 0 \ec
\end{align}
and it should include the mode $z^{d-\D}$ with unit coefficient (this is the Fourier-transform of the mode $\delta(x) z^{d-\D}$).
The solution we are looking for is
\begin{align}
  K_\Delta(k,z) = \frac{2^{1+\nu}}{\Gamma(-\nu)} |k|^{-\nu}
  z^{d/2} \cK_\nu(|k|z) \ed
  \label{KDmom}
\end{align}
Here $\cK_\nu$ is the modified Bessel function, which was chosen because it leads to a regular solution as we take $z \to \infty$.
Expanding the solution at small $z$, we have
\begin{align}
  K_\D(k,z) = 
  \frac{\Gamma(\nu)}{\Gamma(-\nu)}
  \left( \frac{k}{2} \right)^{-2\nu} 
  z^{\D} \left[ 1 + O(z^2) \right]
  + z^{d - \D} \left[ 1 + O(z^2) \right] \ed
  \label{KDkexp}
\end{align}
We see that the $z^{d-\D}$ mode has the required coefficient. 

We can now Fourier transform back to position space, and verify that the propagator takes the known form \cite{Witten:1998qj}
\begin{align}
  K_\D(x,z;x') = \frac{\Gamma(\D)}{\pi^{d/2} \Gamma(-\nu)}
  \frac{z^\D}{ \left( z^2 + |x-x'|^2 \right)^\D } \ed
  \label{KDpos}
\end{align}
Here we will {\it assume} that the functional form of the propagator is as indicated, and will only verify that the coefficient is correct for any $\D \ne d/2$.
To do this, let us set $x'=0$ and isolate the $z^\D$ mode of the propagator. 
The Fourier transform of this mode should be equal to the $z^\D$ mode of the expansion \eqref{KDkexp}. 
This can be verified using the relation\footnote{
For $\D < d/2$ this integral converges and is straightforward to compute.
For $\D > d/2$ it diverges; it can be computed by placing a cutoff in the radial direction, and minimally subtracting the diverging part.
When $\D \ge (d/2) + 1$ additional divergences appear, and we do not consider this case.
}
\begin{align}
  \int d^dx e^{-ik\cdot x} \frac{1}{|x|^{2\D}} =
  \frac{\pi^{d/2} \Gamma(\nu)}{\Gamma(\D)}
  \left( \frac{k}{2} \right)^{-2\nu} \ec
\end{align}
which is valid for $(d/2) - 1 < \D < (d/2) + 1$, $\D \ne d/2$.

\subsubsection{On-shell Action}

Having computed the boundary-to-bulk propagator, we now proceed with the computation of the on-shell action.
We start with the action \eqref{Sfree}, and regulate the theory by placing a cutoff at $z=\e$.
After integrating by parts and using the equations of motion, we find the regularized on-shell action
\begin{align}
  - \frac{1}{2} \int_{z=\e} d^dx \sqrt{\g} 
  \e (\dho_z \phi) \phi + S_{\rm ct} \ec
  \label{SOS3}
\end{align}
where $\g$ is the induced metric on the boundary, and $\sqrt{\g} = \e^{-d}$.
The minus sign in the first term is due to the fact that $z=\e$ is the lower limit of the $dz$ integral.
The first term in this action diverges as we take $\e \to 0$; this can be seen by using the expansion \eqref{phifree}.
We must therefore introduce a counter-term to cancel the divergence.

Let us begin with the case $\D < d/2$, and choose our boundary action to be
\begin{align}
  S_{\rm ct} = \int d^dx \sqrt{\g} \left[
  \frac{\D}{2} \phi^2(x) + 2\nu \e^{d-\D} \phi(x) J(x)
  \right] \ed
  \label{Sct1}
\end{align}
The first term in the boundary action is a counter-term that cancels the divergence in \eqref{SOS3}.
Since we are interested in computing correlators, we would like to impose the boundary condition $\beta(x) = J(x)$ for a source $J(x)$, and we must make sure that our action is compatible with this condition.
This is achieved by the second term in the boundary action \eqref{Sct1}.
Indeed, for $\D < d/2$ the scalar mode $\beta(x) z^{d-\D}$ is not the leading mode in $z$. 
Therefore, we need to impose Neumann-like boundary conditions, in which the variation of the field is left arbitrary.
The variation of the action \eqref{Sfree} with the choice \eqref{Sct1} includes the boundary term
\begin{align}
  \int_{z=\e} d^dx \sqrt{\g} \delta\phi 
  (\D \phi - \e \dho_z \phi + 2\nu \e^{d-\D} J) =
  -2\nu \e^{-\D} \int d^dx \delta\phi [\beta(x) - J(x)] + \cdots
  \ec
\end{align}
where the remaining terms are subleading in $\e$.
As promised, this variation is compatible with the boundary condition.  
The renormalized on-shell action is now
\begin{align}
  S_{\rm on-shell} = \lim_{\e \to 0} \frac{1}{2} \int_{z=\e}
  d^dx \sqrt{h} \phi (\D \phi - \e \dho_z \phi + 4\nu\e^{d-\D}J) =
  \nu \int d^dx \alpha(x) \beta(x) \ed
  \label{SOS0}
\end{align}
Here we used the expansion \eqref{phifree} and the boundary condition $\beta = J$.

Next, consider the case $\D > d/2$.
Here we choose our boundary action to be
\begin{align}
  S_{\rm ct} = \frac{d-\D}{2} \int_{z=\e} d^dx \sqrt{h} \phi^2 \ec
\end{align}
in order to cancel the divergence coming from the bulk action.
In this case the source mode $\beta(x) z^{d-\D}$ is the leading mode near the boundary, so we apply the Dirichlet-like boundary condition $\lim_{z \to 0} z^{\D-d} \delta \phi(x,z) = 0$.
This is compatible with setting $\beta(x) = J(x)$ for any $J(x)$, and there is no need to introduce any additional boundary terms.
The renormalized on-shell action in this case is
\begin{align}
  S_{\rm on-shell} = \lim_{\e \to 0} \frac{1}{2} \int_{z=\e}
  d^dx \sqrt{h} \phi [(d-\D) \phi - \e \dho_z \phi] =
  \nu \int \! d^dx \, \alpha(x) \beta(x) \ed
\end{align}
This is simply the continuation of the action \eqref{SOS0} to the range $\D > d/2$.
We therefore find that the on-shell action for any $\D \ne d/2$ is given by
\begin{align}
  \nu \int \! d^dx \, \alpha(x) \beta(x) \ed
  \label{SOS1}
\end{align}

Let us derive 1-point and 2-point functions of $\cO$ from the on-shell action.
Using the relation
\begin{align}
  \alpha(x) = \frac{\Gamma(\D)}{\pi^{d/2} \Gamma(-\nu)}
  \int d^dx' \frac{\beta(x')}{|x-x'|^{2\D}} \ec
\end{align}
which follows from \eqref{KDpos} and \eqref{phifree}, the renormalized on-shell action becomes
\begin{align}
  S_{\rm on-shell} = \frac{\nu \Gamma(\D)}{\pi^{d/2} \Gamma(-\nu)}
  \int d^dx d^dx' \frac{\beta(x) \beta(x')}{|x-x'|^{2\D}} \ed
\end{align}
The 1-point function is given by (c.f. \eqref{dict})
\begin{align}
  \langle \cO(x) \rangle = - \frac{\delta S_{\rm on-shell}}{\delta \beta(x)}
  = - \frac{2\nu \Gamma(\D)}{\pi^{d/2} \Gamma(-\nu)}
  \int d^dx' \frac{\beta(x')}{|x-x'|^{2\D}} 
  = -2\nu\alpha(x) \ed
\end{align}
We find that the relation 
\begin{align}
  \langle \cO \rangle = -2\nu\a
  \label{Oa}
\end{align}
holds for all $\D \ne d/2$ \cite{Klebanov:1999tb}.\footnote{To compare with \cite{Klebanov:1999tb} one should flip the sign of $\nu$.}
This relation holds for a specific normalization of $\cO$, which is given by its 2-point function.
Differentiating again, we find that
\begin{align}
  \langle \cO(x) \cO(x') \rangle = 
  - \frac{2\nu \Gamma(\D)}{\pi^{d/2} \Gamma(-\nu)}
  \frac{1}{|x-x'|^{2\D}} \ed
  \label{O2pt}
\end{align}
As a check on these results, notice that this 2-point function is reflection positive for all $\D \ne d/2$.

\subsection{The Case $\D = d/2$}

In this case $\nu=0$, and the near-boundary expansion of the scalar is
\begin{align}
  \phi(x,z) = z^{d/2} \log(z \mu) 
  \left[ \beta(x) + O(z^2) \right]
  + z^{d/2} \left[ \alpha(x) + O(z^2) \right]
  \ed \label{phifreemar}
\end{align}
Here $\b$ corresponds to the source, because one can turn on $\a$ without $\b$ but not the other way around \cite{Klebanov:1999tb}.
The boundary-to-bulk propagator in momentum space is
\begin{align}
  K_\Delta(k,z) = - z^{d/2} \cK_0(k z) \ed
  \label{propmar}
\end{align}
This can be verified by checking that the $z^{d/2} \log(z)$ mode of this solution has unit coefficient.

The on-shell action is given by \eqref{SOS3}, and the counter-term action that renormalizes the theory in this case is
\begin{align}
  S_{\rm ct} = \left( \frac{d}{4} +  
  \frac{1}{2\log(\epsilon \mu)}
  \right)
  \int d^dx \sqrt{h} \phi^2 \ed
\end{align}
This counter-term is introduced to cancel the divergence in the first term in \eqref{SOS3}.
In this case the source term is the leading term near the boundary, so we can impose Dirichlet-like boundary conditions (as we did for $\D > d/2$), which are compatible with the boundary condition $\beta(x) = J(x)$.
After taking $\e \to 0$, we find the renormalized on-shell action
\begin{align}
  S_{\rm on-shell} = \int \! d^dx \, \alpha(x) \beta(x) \ed
  \label{SOSmar}
\end{align}

In this case it is convenient to proceed in momentum space.
Expanding the propagator \eqref{propmar} near the boundary, we see that
\begin{align}
  K_\D(k,z) = z^{d/2} \log(\mu z) [1 + O(z^2)] +
  z^{d/2} \left[ \gamma + \log(k/2\mu) + O(z^2) \right] \ec
\end{align}
where $\gamma$ is Euler's gamma, whose appearance is an artifact of our choice of scale $\mu$.
Comparing with the expansion \eqref{phifreemar} (after Fourier-transforming it), we find the momentum-space relation
\begin{align}
  \alpha(k) = \left[ \gamma + \log(k/2\mu) \right] \beta(k) \ed
\end{align}
Plugging this in the on-shell action \eqref{SOSmar}, and differentiating with respect to $\beta$, we find the relation
\begin{align}
  \langle \cO \rangle = -2 \alpha \ed
  \label{Oamar}
\end{align}
Differentiating again, we find the 2-point function
\begin{align}
  \langle \cO(k) \cO(k') \rangle =
  -2 \log(k/\mu) \delta(k+k') + \cdots \ec
\end{align}
where we have omitted terms that are constant in the momentum, because they correspond to contact terms in position space.
One can now transform back to position space using the relation
\begin{align}
  \int \! d^dx \, e^{-ik \cdot x} \frac{1}{|x|^d} =
  - \frac{2\pi^{d/2}}{\Gamma(d/2)} \log(k\varepsilon) + \mathrm{const} \ec
\end{align}
where $\varepsilon$ is a cutoff in position space ($|x| > \varepsilon$).
At separated points, we find that
\begin{align}
  \langle \cO(x) \cO(x') \rangle = \frac{\Gamma(d/2)}{\pi^{d/2}}
  \frac{1}{|x-x'|^d} \ed
  \label{O2ptmar}
\end{align}
This determines the normalization of the operator $\cO$ in our calculation, and also shows that this procedure leads to a reflection-positive 2-point function.

\section{Solving the Scalar Equation of Motion}
\label{soleq}

In this section we solve the equation of motion \eqref{bulkeq} for the scalar field near the boundary, to leading order in $\eta$.
The equation is
\begin{align}
 z^2 \phi'' + z^2 \square_x \phi + (1-d) z \phi' - m^2 \phi = 
 \theeta \phi^{n-1}(x,z)
 \ecq n \ge 3 \ec
\end{align}
where $\phi' = \dho_z \phi$.
Let us expand the solution as
\begin{align}
  \phi = \phi_0 + \theeta \phi_1 + O(\theeta^2) \ec
\end{align}
where the free field solution is given by 
\begin{align}
  \phi_0 = z^\D \left[ \a(x) + O(z^2) \right] + 
  z^{d-\D} \left[ \b(x) + O(z^2) \right]
  \ed
\end{align}
The equation for $\phi_1$ is
\begin{align}
 z^2 \phi_1'' + z^2 \square_x \phi_1 +
 (1-d) z \phi_1' - m^2 \phi_1 
 &= \phi_0^{n-1}(x,z) \cr
 &= \alpha^{n-1}(x) z^{d-\D} + O(z^{d-\D+2\nu})
 \ed \label{eq2solve}
\end{align}
The effect of the $z^2 \square_x \phi_1$ term is subleading in $z$, and the solution is
\begin{align}
  \phi_1 = \frac{1}{2\nu} \alpha^{n-1}(x) z^{d-\D} \log(\mu z)
  + O(z^{d-\D+2\nu}) \ed
\end{align}
The full solution near the boundary is thus
\begin{align}
  \phi(x,z) &= z^\D \left[ \a(x) + O(z^2) \right] + 
  z^{d-\D} \left[ \b(x) + O(z^2) \right]
  \cr &\quad
  + \frac{\eta}{2\nu} \alpha^{n-1}(x) z^{d-\D} \log(\mu z)
  + O(\eta z^{d-\D+2\nu}) + O(\eta^2)
  \ed
\end{align}

\end{document}